# Minimum Component Based First-Order Inverting and Non-inverting Outputs of All-Pass Filter at the Same Circuit

J. Mohan[1], S. Maheshwari[2], and D. S. Chauhan[3]
[1]Department of Electronics and Communications, Jaypee University of Information Technology,
Waknaghat, Solan-173215 (India)
Email:jitendramv2000@rediffmail.com
[2]Department of Electronics Engineering, Z. H. College of Engineering and Technology,
Aligarh Muslim University, Aligarh-202002 (India)
Email:Sudhanshu_maheshwari@rediffmail.com
[3]Department of Electrical Engineering, Institute of Technology, Banaras Hindu University,
Varanasi-221005 (India)
Email:pdschauhan@gmail.com

*Abstract*— **In this paper, a new voltage-mode first order all-pass filter using minimum active and passive components is presented. The proposed circuit employs one fully differential second generation current conveyor (FDCCII), one grounded capacitor, one resistor and offers the following advantages: the use of only grounded capacitor which is attractive for integrated circuit implementation, low active and passive sensitivities, providing inverting and non-inverting voltage-mode all-pass responses simultaneously from the single circuit and no requirement for component matching conditions. The theory is validated through PSPICE simulation using TSMC 0.35μm CMOS process parameters.**

*Index Terms*—**first term, second term, third term, fourth term, fifth term, sixth term**

## I. Introduction

First order all-pass filters are an important class of analogue signal processing circuits which have been extensively researched in the technical literature [1-2] due to their utility in communication and instrumentation systems, for instance as a phase equalizer, phase shifter or for realizing quadrature oscillators band pass filters etc.

A voltage-mode all-pass filter with minimum component is expected to use two passive components and one active element. Since two component based circuits are free from matching problems as both pole and zero frequency depend on the same two components such circuits would benefit from easy control over the pole frequency, as only a single element need to be controlled unlike the circuits require three components. In the literature, several voltage-mode all-pass filter circuit employing different types of active elements such as current conveyors and its different variations have been reported [3-20]. Some voltage-mode circuits benefit from minimum component feature and hence require no matching constraints [ ]. However, none of the reported circuit using minimum component count provides inverting and non-inverting all-pass response simultaneously from the single circuit.

This paper proposes a new circuit for realizing inverting and non-inverting voltage-mode all-pass filters with different phase responses together using single active element, and two passive components. This feature not collectively exhibited in any of the reported work in the literature, including the most recent and useful circuit [19-20]. The proposed circuit is based on FDCCII, an active element to improve the dynamic range in mixed mode application, where fully differential signal processing is required [21]. PSPICE simulation results using TSMC 0.35μm CMOS parameters are given to validate the circuits.

## II. Circuit Description

The fully differential second generation current conveyor (FDCCII) is an eight terminal analog building block with the defining matrix equation of the form

$$\begin{bmatrix} I_{Y1} \\ I_{Y2} \\ I_{Y3} \\ I_{Y4} \\ V_{X+} \\ V_{X-} \\ I_{Z+} \\ I_{Z-} \end{bmatrix} = \begin{bmatrix} 0 & 0 & 0 & 0 & 0 & 0 & 0 & 0 \\ 0 & 0 & 0 & 0 & 0 & 0 & 0 & 0 \\ 0 & 0 & 0 & 0 & 0 & 0 & 0 & 0 \\ 0 & 0 & 0 & 0 & 0 & 0 & 0 & 0 \\ 1 & -1 & 1 & 0 & 0 & 0 & 0 & 0 \\ -1 & 1 & 0 & 1 & 0 & 0 & 0 & 0 \\ 0 & 0 & 0 & 0 & 1 & 0 & 0 & 0 \\ 0 & 0 & 0 & 0 & 0 & -1 & 0 & 0 \end{bmatrix} \begin{bmatrix} V_{Y1} \\ V_{Y2} \\ V_{Y3} \\ V_{Y4} \\ I_{X+} \\ I_{X-} \\ V_{Z+} \\ V_{Z-} \end{bmatrix} \quad (1)$$

The CMOS implementation of FDCCII is shown in Fig. 1 [21]. FDCCII is a useful and versatile active element for analog signal processing [21-22].

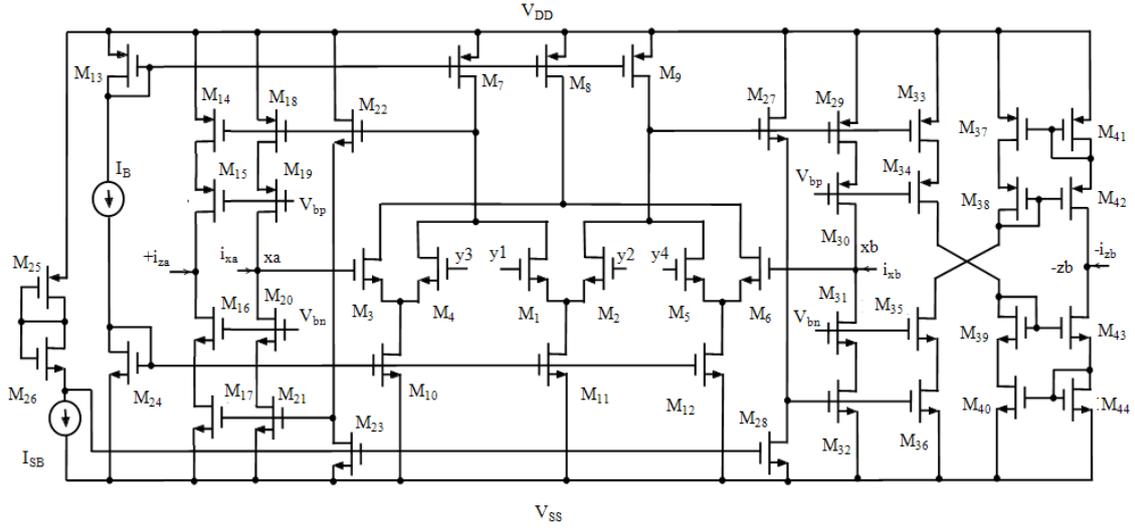

Figure 1. Fully differential second generation current conveyor CMOS implementation

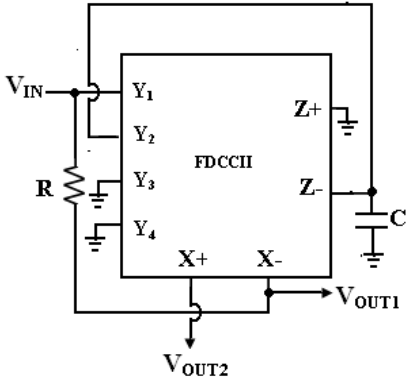

Figure 2. Proposed voltage-mode all-pass filter circuit.

The proposed voltage-mode all-pass filter circuit using a single FDCCII, one grounded capacitor and one resistor is shown in Fig. 2. The circuit is characterized by the following voltage transfer functions

$$\frac{V_{OUT1}}{V_{IN}} = -\frac{s - (1/RC)}{s + (1/RC)} \quad (2)$$

$$\frac{V_{OUT2}}{V_{IN}} = \frac{s - (1/RC)}{s + (1/RC)} \quad (3)$$

Equations (2)-(3) is the standard first-order all-pass transfer function. The circuits of Fig. 2, thus provides a unity gain at all frequencies and frequency dependent phase function ($\Phi$) with a value $\Phi = -2\tan^{-1}(\omega RC)$ for equation (2) and $\Phi = 180^0 - 2\tan^{-1}(\omega RC)$ for equation (3).

The salient features of the proposed circuit is realizing inverting and non-inverting voltage-mode all-pass filters with different phase responses simultaneously using single active element, and two passive components, the feature not exhibited together in any of the available works [3-20]. The circuit also enjoys one of the voltage output at low impedance thus making them suitable for voltage-mode cascading.

### III. NON-IDEAL ANALYSIS

To account for non ideal sources, two parameter $\alpha$ and $\beta$ are introduced where $\alpha_i$ (i=1,2) accounts for current transfer gains and $\beta_i$ (i=1,2,3,4,5,6) accounts for voltage transfer gains of the FDCCII. These transfer gains differ from unity by the voltage and current tracking errors of the FDCCII. More specifically, $\alpha_i = 1-\delta_i$, ($|\delta_i| \ll 1$) where current tracking error $\delta_1$ (from X+ to Z+) and $\delta_2$ (from X- to Z-). Similarly, $\beta_i = 1-\varepsilon_i$, ($|\varepsilon_i| \ll 1$) where voltage tracking errors $\varepsilon_1$ (from $Y_1$ to X+), $\varepsilon_2$ (from $Y_2$ to X+), $\varepsilon_3$ (from $Y_3$ to X+), $\varepsilon_4$ (from $Y_1$ to X-), $\varepsilon_5$ (from $Y_2$ to X-), and $\varepsilon_6$ (from $Y_4$ to X-), respectively. Incorporating the two sources of error onto ideal input-output matrix relationship of the modified FDCCII leads to:

$$\begin{bmatrix} I_{Y1} \\ I_{Y2} \\ I_{Y3} \\ I_{Y4} \\ V_{X+} \\ V_{X-} \\ I_{Z+} \\ I_{Z-} \end{bmatrix} = \begin{bmatrix} 0 & 0 & 0 & 0 & 0 & 0 & 0 & 0 \\ 0 & 0 & 0 & 0 & 0 & 0 & 0 & 0 \\ 0 & 0 & 0 & 0 & 0 & 0 & 0 & 0 \\ 0 & 0 & 0 & 0 & 0 & 0 & 0 & 0 \\ \beta_1 & -\beta_2 & \beta_3 & 0 & 0 & 0 & 0 & 0 \\ -\beta_4 & \beta_5 & 0 & \beta_6 & 0 & 0 & 0 & 0 \\ 0 & 0 & 0 & 0 & \alpha_1 & 0 & 0 & 0 \\ 0 & 0 & 0 & 0 & 0 & -\alpha_2 & 0 & 0 \end{bmatrix} \begin{bmatrix} V_{Y1} \\ V_{Y2} \\ V_{Y3} \\ V_{Y4} \\ I_{X+} \\ I_{X-} \\ V_{Z+} \\ V_{Z-} \end{bmatrix} \quad (4)$$

The circuits of Fig. 2 is analyzed using (4) and the non-ideal voltage transfer functions are found as

$$\frac{V_{OUT1}}{V_{IN}} = -\beta_4 \left( \frac{s - \beta_5 \alpha_2 / (RC)}{s + \beta_5 \alpha_2 / (RC)} \right) \quad (5)$$

$$\frac{V_{OUT2}}{V_{IN}} = \beta_1 \left( \frac{s + \alpha_2(\beta_1\beta_5 - \beta_2 - \beta_2\beta_4)/(\beta_1 RC)}{s + \beta_5\alpha_2/(RC)} \right) \quad (6)$$

Thus, the pole frequency ($\omega_o$) of the first order all-pass filter circuit of Fig. 2 can be expressed as

$$\omega_o = \frac{\beta_5 \alpha_2}{RC} \quad (7)$$

From (7), the pole frequency ($\omega_o$) sensitivities can be expressed as

$$S_{R,C}^{\omega_o} = -1;\ S_{\beta_5,\alpha_2}^{\omega_o} = 1;\ S_{\beta_1,\beta_2,\beta_3,\beta_4,\beta_6,\alpha_1}^{\omega_o} = 0; \quad (8)$$

From (8), the sensitivities of active and passive components with respect to pole frequency ($\omega_o$) are within unity in magnitude. Thus, the circuit enjoys attractive active and passive sensitivity performance.

## IV. SIMULATION RESULTS

The proposed circuits were verified using PSPICE simulation. The FDCCII was realized using CMOS implementation as shown in Fig. 1 [21] and simulated using TSMC 0.35μm, Level 3 MOSFET parameters as listed in Table 2. The aspect ratio of the MOS transistors are listed in Table 3, with the following DC biasing levels $V_{dd} = -V_{ss} = 3V$, $V_{bp} = V_{bn} = 0V$, and $I_B = I_{SB} = 1.2mA$.

**Table 2.** 0.35μm level 3 MOSFET parameters

| NMOS: |
| --- |
| LEVEL=3 TOX=7.9E-9 NSUB=1E17 |
| GAMMA=0.5827871 PHI=0.7 VTO=0.5445549 |
| DELTA=0 UO=436.256147 ETA=0 THETA=0.1749684 |
| KP=2.055786E-4 VMAX=8.309444E4 |
| KAPPA=0.2574081 RSH=0.0559398 NFS=1E12 TPG=1 |
| XJ=3E-7 LD=3.162278E-11 WD=7.04672E-8 |
| CGDO=2.82E-10 CGSO=2.82E-10 CGBO=1E-10 |
| CJ=1E-3 PB=0.9758533 MJ=0.3448504 |
| CJSW=3.777852E-10 MJSW=0.3508721 |
| PMOS: |
| LEVEL =3 TOX = 7.9E-9 NSUB=1E17 |
| GAMMA=0.4083894 PHI=0.7 VTO=-0.7140674 |
| DELTA=0 UO=212.2319801 ETA=9.999762E-4 |
| THETA=0.2020774 KP=6.733755E-5 |
| VMAX=1.181551E5 KAPPA=1.5 RSH=30.0712458 |
| NFS=1E12 TPG=-1 XJ=2E-7 LD=5.000001E-13 |
| WD=1.249872E-7 CGDO=3.09E-10 CGSO=3.09E-10 |
| CGBO=1E-10 CJ=1.419508E-3 PB=0.8152753 MJ=0.5 |
| CJSW=4.813504E-10 MJSW=0.5 |

**Table 3.** Transistor aspect ratios for the circuit shown in Fig. 1

| Transistors | W(μm) | L(μm) |
| --- | --- | --- |
| M1-M6 | 60 | 4.8 |
| M7-M9, M13 | 480 | 4.8 |
| M10-M12, M24 | 120 | 4.8 |
| M14, M15, M18, M19, M25, M29, M30, M33, M34 | 240 | 2.4 |
| M16, M17, M20, M21, M26, M31, M32, M35, M36 | 60 | 2.4 |
| M22, M23, M27, M28 | 4.8 | 4.8 |

The circuit of Fig. 2 was designed with C=1 nF and R=1 kΩ. The designed pole frequency was 159.2 KHz. The phase and gain plot is shown in Fig. 3. The phase is found to vary with frequency from 180° to 0 for $V_{OUT1}$ and from 0 to -180° for $V_{OUT2}$ with a value of 90° and -90° at the pole frequency, and the pole frequency was found to be 155.6 KHz, which is in error by ≈2% with the designed value. The circuit was next used as a phase shifter introducing 90° and -90° shift to a sinusoidal voltage input of 1volt peak at 155.6 KHz was applied. The input and ±90° phase shifted output waveforms (given in Fig. 4) which verify circuit as a phase shifter. The THD variation at the output for varying signal amplitude at 155.6 KHz was also studied and the results shown in Fig. 5. The THD for a wide signal amplitude (few mV-1000mV) variation is found within 4.3% at 155.6 KHz.

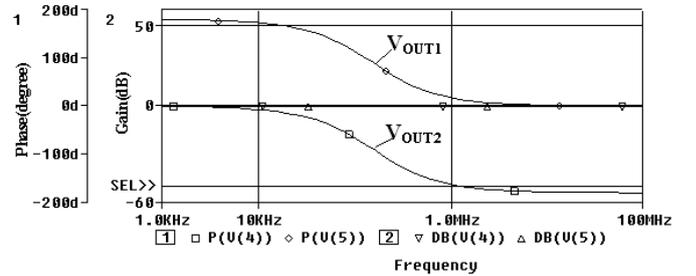

Figure 3. Simulated gain and phase response of Fig. 2 at voltage output terminals.

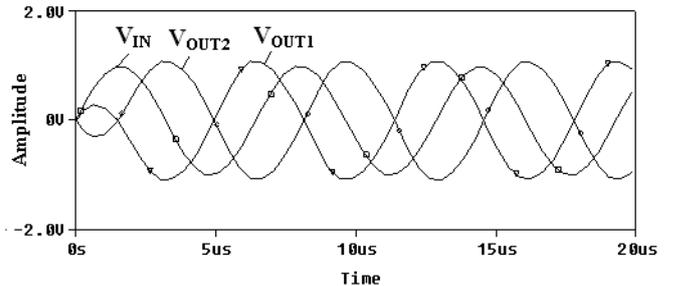

Figure 4. Time-domain waveforms of the Fig. 2 at input frequency 155.6 KHz

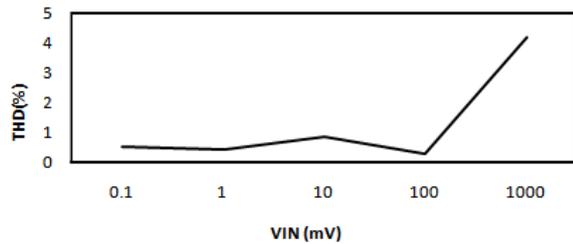

Figure 5. THD variation at output with signal amplitude at 155.6 KHz

## V. INTEGRATION ASPECTS

The proposed circuit is can be conveniently implemented in CMOS technology. The resistor can be replaced by active-MOS resistor with added advantage of tunability through external voltage [23]. Similarly, there are techniques of implementing capacitor in MOS technology [24]. Since the used capacitor is in grounded form, it is further favourable as far as implementation is concerned. Thus the proposed circuits are quite suitable for IC implementation.

## VI. CONCLUSION

This paper has presented a new voltage-mode first order all-pass filter employing one FDCCII, one grounded capacitor and resistor. The salient features of the proposed circuit is use of grounded capacitor, suitable for IC implementation, providing inverting and non-inverting voltage-mode all-pass responses simultaneously from the circuit and also providing one of the voltage output at low impedance thus making them suitable for voltage-mode cascading. The proposed circuit is verified through PSPICE simulation using 0.35μm TSMC parameters. The integration of the proposed circuit is an open area for further research.